\documentclass[aps,reprint,superscriptaddress,nofootinbib
]{revtex4-1}

\usepackage{graphicx}%
\usepackage{dcolumn}%
\usepackage{bm}%
\usepackage{xspace}%
\usepackage{natbib}
\usepackage{lipsum}
\usepackage{amsmath}
\usepackage{upgreek}
\usepackage{float}
\usepackage{epsf}
\usepackage{amsfonts}
\usepackage{amssymb}
\usepackage{times}
\usepackage{hyphenat}

\newcommand{\DNatracer}{\mbox{$D^{\rm \ast}_{\rm Na}$}\xspace}
\newcommand{\DKtracer}{\mbox{$D^{\rm \ast}_{\rm K}$}\xspace}

\newcommand{\wNa}{\mbox{$w_{\rm Na}$}\xspace}
\newcommand{\wK}{\mbox{$w_{\rm K}$}\xspace}

\newcommand{\pc}{\mbox{$p_{\rm c}$}\xspace}

\newcommand{\NaI}{\mbox{${\rm Na}_{\rm I}$}\xspace}
\newcommand{\KI}{\mbox{${\rm K}_{\rm I}$}\xspace}

\begin{document}

\title{Potassium self-diffusion in a K-rich single-crystal alkali feldspar}



\author{F. Hergem\"oller}
\email{fabian.hergemoeller@wwu.de}
\affiliation{Institut f\"ur Materialphysik, University
of M\"unster, Wilhelm-Klemm-Str. 10, 48149 M\"unster, Germany}
\author{M. Wegner}
\affiliation{Institut f\"ur Materialphysik, University
of M\"unster, Wilhelm-Klemm-Str. 10, 48149 M\"unster, Germany}
\author{M. Deicher}
\affiliation{Technische Physik, Universit\"at des Saarlandes , 66123 Saarbr\"ucken, Germany}
\author{H. Wolf}
\affiliation{Technische Physik, Universit\"at des Saarlandes , 66123 Saarbr\"ucken, Germany}
\author{F. Brenner}
\affiliation{ Institute for Chemical Technologies and Analytics, Vienna University of Technology,
Getreidemarkt 9/164 AC, 1060 Vienna, Austria}
\author{H. Hutter}
\affiliation{ Institute for Chemical Technologies and Analytics, Vienna University of Technology,
Getreidemarkt 9/164 AC, 1060 Vienna, Austria}
\author{R. Abart}
\affiliation{Department of Lithospheric Research, University of Vienna,
Althanstr. 14, 1090 Vienna, Austria}
\author{N. A. Stolwijk}
\email{stolwij@wwu.de}
\affiliation{Institut f\"ur Materialphysik, University
of M\"unster, Wilhelm-Klemm-Str. 10, 48149 M\"unster, Germany}



\begin{abstract}
The paper reports potassium diffusion measurements performed on gem quality single-crystal alkali feldspar in the temperature range from 1169 to \mbox{1021 K}. Natural sanidine from Volkesfeld, Germany was implanted with ${}^{43}\rm{K}$ at the ISOLDE/CERN radioactive ion-beam facility normal to the $\left(001\right)$ crystallographic plane. Diffusion coefficients are well described by the Arrhenius equation with an activation energy of $2.4 \, \, \rm{eV}$ and a pre-exponential factor of $5 \times 10^{-6} \, \, \rm{m}^2/\rm{s}$, which is more than three orders-of-magnitude lower than the ${}^{22}\rm{Na}$ diffusivity in the same feldspar and the same crystallographic direction. State-of-the-art considerations including ionic conductivity data on the same crystal and Monte Carlo simulations of diffusion in random binary alloy structures point to a correlated motion of K and Na through the interstitialcy mechanism.
\end{abstract}

\pacs{6172Ji, 6630Dn, 6630Hs}

\maketitle
\section{\label{sect:intro}Introduction}
The rock-forming alkali feldspars belong to the most abundant minerals in the Earth's crust and are formed as a solid solution between the sodium ($\rm{NaAlSi}_{3}\rm{O}_{8}$, albite) and potassium ($\rm{KAlSi}_{3}\rm{O}_{8}$, orthoclase) end-member compositions. Well-founded knowledge of self-diffusion data in alkali feldspar is a prerequisite for interpreting existing \emph{inter}diffusion data \cite{petrishcheva14, schaeffer14} that, in turn control re-equilibration features in alkali feldspar that pertain to evolution and dynamics of the crust \cite{Abart09}. Until now, most studies on alkali diffusion in alkali feldspar concern the sodium component because of the availability of the radioisotope ${}^{22}\rm{Na}$ with a half-life of $t_{1/2} = 2.6 \,\, \rm{a}$ and its suitability for the radiotracer method (see e.g. review by Cherniak \cite{cherniak2010}). Potassium self-diffusion has hitherto been investigated by Lin and Yund \cite{Lin1972} by using the long-lived radioisotope ${}^{40}\rm{K}$ $\left( t_{1/2} = 1.3 \times 10^9 \,\, \rm{a} \right)$ and by Foland \cite{foland74} by using the stable isotope ${}^{41}\rm{K}$ (natural abundance of 6.7 \%). In both studies a bulk-exchange method was employed to investigate grains from crushed feldspar. An advantage of this method is that there is no need for large and almost perfect single-crystal alkali feldspar samples, which are rare in nature and cannot be synthesized. However, this method is unable to provide any information about tracer depth distributions and suitable diffusion models must therefore be presumed, making it non-sensitive to inherent features such as diffusion anisotropy or structural inhomogeneities. Moreover, these minerals have not been investigated by impedance spectroscopy and therefore these diffusion coefficients cannot be compared to ionic conductivity data. 

The diversity of investigated alkali feldspars and applied methods has made it difficult to reliably link existing studies with each other and to discuss the underlying mechanisms of alkali diffusion based on sound arguments. So far, it was found that activation energies of K and Na self-diffusion clearly differ so that the ratio of diffusion coefficients $\DNatracer / \DKtracer$ is approximately $1000 / 1$ at $1000 \,\, {}^{\circ}\rm{C}$ \cite{Behrens1991} for the K-rich orthoclase and approximately $600 / 1$ at $800 \,\, {}^{\circ}\rm{C}$ \cite{Kasper75} for the Na-rich albite. Frenkel pairs are likely to be the major point defects in the feldspar structure because the high Si-O bonding energies make Schottky defect formation extremely unfavourable \cite{Behrens1991}. In general, the vacancy mechanism and the interstitialcy mechanism should therefore be considered as predominant in alkali diffusion processes. In the latter mechanism, an interstitial $\NaI$ or $\KI$ atom moves by ''pushing'' a substitutional alkali ion to a neighbouring interstitial site. A further crucial diffusion mechanism is based on direct interstitial jumps of an interstitial ion to a neighbouring interstitial site without exchanging with substitutional atoms of the alkali sublattice.

Self-diffusion of atoms in binary (Na, K sublattice) or multi-component systems is generally affected by correlation effects among individual atomic movements. The jump direction of a labelled self-atom (tracer) depends on the relative position of the point defect next to it. After a first exchange, the atom is still in direct vicinity to the point defect and therefore has an increased probability for a consecutive jump in reversed direction. This \emph{geometric} correlation effect is influenced by the diffusion mechanism as well as the pertaining crystal structure and can be expressed by correlation factors $f$ in the range $\frac{1}{3} \leq f \leq 1$ \cite{book-mehrer}. In addition, the distribution of Na and K atoms in the environment of a tracer atom causes a \emph{physical} correlation effect that arises from differences in the temperature dependent jump frequencies $\left(\wK / \wNa \ll 1\right)$ of these components. Influenced by chemical composition and the diffusion mechanism this correlation is potentially strong since both components share the same sublattice. Below a certain site fraction of Na, commonly referred to as percolation threshold $\pc$, diffusion pathways of the faster moving ions (Na) lose their percolation ability and the Na diffusion coefficient becomes dependent on the jump frequencies of the slower ions (K). A quantitative analysis of correlation effects (e.g. through Monte Carlo simulations) is an approach to unravel the underlying diffusion mechanisms of alkali diffusion in alkali feldspar, but requires the most precise experimental results available.

The present results on potassium self-diffusion are based on direction-sensitive serial sectioning of gem-quality single-crystal alkali feldspar and the use of ${}^{43}\rm{K}$ $\left( t_{1/2} = 22.3 \,\, \rm{h} \right)$ as radiotracer.
A natural sanidine from Volkesfeld (VF), Germany that was also used in our previous studies to investigate Na self-diffusion \cite{wilangowski15} and the ionic conductivity \cite{hamid15} in the same crystallographic direction, i.e., normal to $\left( 001\right)$, was selected as sample material. This allows us to compare alkali transport properties with ionic conductivity in a reliable way. Basic studies concerning correlation effects in randomly ordered binary alloys based on the vacancy mechanism \cite{wilan-philmag15} and the interstitialcy mechanism \cite{wilan-jcp, wilangowski16} will be used to discuss the mechanisms of alkali diffusion in alkali feldspar.

\section{\label{sect:exp}Experimental procedure}
\subsection{\label{sect:sample}Sample material }

In the present study, the diffusion of ${}^{43}\rm{K}$ in the framework silicate alkali feldspar $\left(\rm K,Na \right)\left[\rm Al\rm Si_{3}\rm O_{8}\right]$ was investigated by the radiotracer method. The material is a natural sanidine feldspar from the Eifel, Germany. For alkali feldspar from this provenance the detailed chemical analysis has been given by Hofmeister et al. \cite{Hofmeister1984} and by Demtr\"oder \cite{demtroeder11} based on electron microprobe analyses (EMPA). The chemical composition of the alkali sublattice is of special interest when K and Na self-diffusion and the diffusion correlation effects pertaining to this are interpreted. Based on the EMPA data the K to cation ratio is approximately $C_{\rm{K}} = c_{\rm{K}} /\left(c_{\rm{Na}}+c_{\rm{K}} +c_{\rm{Ba}} +c_{\rm{Fe}}\right) = 0.83$ (and accordingly $C_{\rm{Na}} = 0.15$) , where $C$ is a site fraction of atoms and $c$ is a concentration in atoms per volume. 

The sample pieces where cleaved from a large single crystal along the $\left( 001\right)$ cleavage plane, cut to a size of approximately $8 \, \,\rm mm$ in diameter and $2 \,\, \rm mm$ in height, and finally polished. The samples did not intendedly undergo any thermal treatment\footnote{ Freer et al. \cite{Freer97} reported an unusual fast Al-Si exchange of the VF feldspar compared to other alkali feldspars when heated dry. This effect disappears when the VF feldspar is heated at elevated temperatures $\geq 1050 \,\, {}^\circ \rm C$ for several days. In the present study we refrained from any thermal pre-treatment in order to investigate the VF feldspar under its natural condition, just as in our previous studies.} before the ${}^{43}\rm{K}$ implantations were carried out. After completion of the radiotracer experiments the samples were examined by energy-dispersive X-ray spectroscopy (EDS) for signs of Na-K interdiffusion. Since only a limited volume of the samples was destructed by radiotracer depth-profiling, the near-surface chemistry could be measured and compared to the chemistry at the floor of the etch crater. The resulting EDS maps showed no traceable changes in Na and K concentrations, which indicates that the self-diffusion experiments were not influenced by interdiffusion of these components with the ambience. 

For further validation, the sample that was annealed at the lowest temperature, i. e. $1021 \,\, \rm{K}$, was subjected to depth-profiling by secondary ion mass spectrometry (ToF-SIMS) at the TU Vienna. This measurement showed no considerable changes in sample chemistry along the investigated depth range. However, ToF-SIMS measurements of the $1173 \,\, \rm{K}$ sample revealed a considerable depletion of Na towards the sample surface whereas K, Al and Si signals remained almost constant over the investigated depth range. By further EDS measurements it could be demonstrated that this observation is probably related to the fact that the investigated spot was in direct contact to a Macor ceramic sample holder during a first diffusion experiment using the opposite side of the same sample, and hence this finding does not relate to the sample chemistry within the volume that was examined by the radiotracer experiments.

\subsection{\label{sect:radiotracer}Implantation and radiotracer experiments}
\begin{figure}
 \includegraphics[width=1.0\columnwidth ,clip]{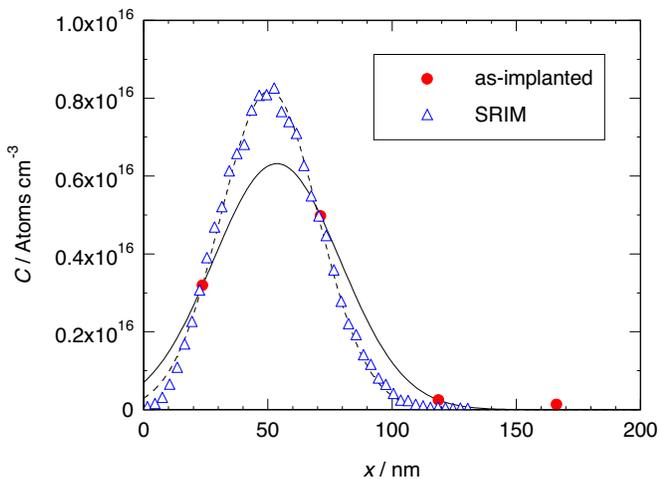}
\caption{Concentration of implanted ${}^{43}\rm K$ as a function of depth $x$ obtained by serial sectioning (circles) and by SRIM simulations (triangles) for an implantation energy of $50 \,\, \rm keV$. A Gaussian function (\ref{eq:gauss}) was fitted to the data points (lines) to determine the mean ion range $x_{0}$ and implantation width $\sigma$. The area of both distributions was normalized to the ion dose of $4 \times 10^{11} \,\, \rm atoms \,\, cm^{-2}$. }
\label{fig:1}
\end{figure}

 Implantations of ${}^{43}\rm{K}$ with an energy of $50 \,\, \rm{keV}$ were carried out at the radioactive ion-beam facility ISOLDE/CERN with typical doses of $4.0 \times 10^{11} \,\, \rm atoms \,\, cm^{-2}$. The ion beam was collimated to a size of $5 \times 5\,\, \rm{mm}$ and $\gamma$-ray spectra showed no contamination with other $\gamma$-emitting isotopes. The corresponding projected range $x_0$ of ${}^{43}\rm{K}$ in alkali feldspar was determined with an implanted, but otherwise thermally untreated control sample (the same method as for the diffusion measurements was applied and is described below). The measured depth distribution is fairly well described by a Gaussian function
\begin{equation}
C \left( x \right) = C_0 \exp{\left(-\frac{\left(x-x_0\right)^2}{2\sigma^2}\right)}
\label{eq:gauss}
\end{equation}
with a peak concentration $C_{0}$ at $x_0 = 54\,\, \rm{nm}$ and a mean width of $\sigma = 26\,\, \rm{nm}$. In figure \ref{fig:1} this result is compared to the calculated ion distribution in feldspar obtained by the SRIM software package (version 2013.00) developed by Biersack and Ziegler \cite{Ziegler2010}. The calculated distribution shows a Gaussian shape and the corresponding values for $x_{0}$ and $\sigma$ (cf. table \ref{tab:table1}) fall closely to those obtained by fitting to the experimental data. The peak concentration $C_{0}$ however slightly differs, but it is used as a free fitting parameter in our analysis presented in section \ref{sect:results}.
After implantation, each sample was annealed at temperatures between 1169 to 1021 K for times between 20 min and 3.2 h in an integral high vacuum diffusion facility that was constructed at the University of Saarbr\"ucken \cite{KronenbergPhd}. Sample temperatures were continuously recorded to determine the effective annealing time (21 to 192 min, cf. table 1) during each experiment. After thermal treatment the samples were sectioned by ion-beam sputtering with a depth resolution of $24$ to $90\,\, \rm{nm}$. Each section was separately collected on a kapton film and transported to a NaI-detector to determine its $\gamma$-ray count rate. The time between implantation and detection was less than three half-lives in all cases. An appropriate short-lived correction \cite{junod1974corrections} for ${}^{43}\rm{K}$ $\left( t_{1/2} = 22.3 \,\, \rm{h} \right)$ was applied to each $\gamma-$ray spectrum to determine the relative tracer concentration. The diffusion profiles, i.e., the relative tracer concentrations as a function of depth $x$, were then obtained by measuring the total profile depth. This was done by mechanical surface profiling of the samples after sputtering.

\section{\label{sect:results} ${}^{43}\rm{K}$ diffusion results}

\begin{figure}
\includegraphics[width=1.0\columnwidth ,clip]{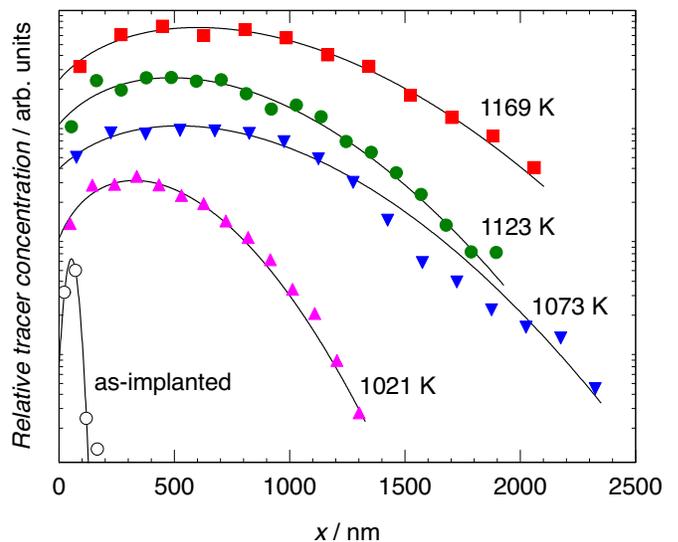}
\caption{Diffusion profiles of ${}^{43}\rm K$ in alkali feldspar normal to $\left( 001 \right)$ for different annealing temperatures $T$. Solid lines represent fittings to the data points according to equation (\ref{eq:strohm}). To enhance visibility of all slopes the relative concentrations are shifted along the ordinate. The ${}^{43}\rm K$ distribution as-implanted is shown for comparison (open circles).}
\label{fig:2}
\end{figure}

\begin{figure}
\includegraphics[width=1.0\columnwidth ,clip]{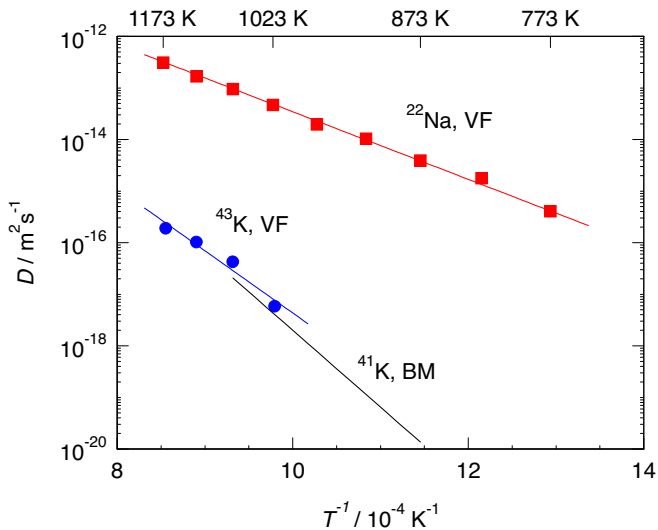}
\caption{Arrhenius plot of the diffusion coefficients of ${}^{43}\rm K$ in alkali feldspar from Volkesfeld (VF), Germany (circles). Corresponding results for the diffusion of ${}^{22}\rm Na$ are included for comparison \cite{wilangowski15} (squares), as well as the Arrhenius relation for the ${}^{41}\rm K$ diffusion in a more K-rich orthoclase from Benson Mines (BM), USA \cite{foland74} (solid line, based on a cylindrical grain model).}
\label{fig:3}
\end{figure}

The results of four ${}^{43}\rm{K}$ depth distributions in alkali feldspar after diffusion annealing are presented in figure \ref{fig:2} together with results for the as-implanted sample (cf. figure \ref{fig:1}). To prevent overlapping and intersecting data, the measured profiles were individually shifted along the ordinate. It can be seen from the figure that the diffusion length is at least ten times larger than the mean implantation depth $x_0$, in all cases. As a consequence, the uncertainty in $x_{0}$ and $\sigma$ (cf. figure \ref{fig:1}) has negligibly effects on the deduced values of $D_{\rm{K}}^{\ast}$. Furthermore, a characteristic decline in tracer concentration towards the surface is observed in all experiments. It seems less likely that this effect is based on a supersaturation of vacancies due to implantation damage, since at the high temperatures employed a virtually immediate equilibration is expected.
A suitable solution of the diffusion equation should converge to the initial tracer distribution described by equation (\ref{eq:gauss}) for negligibly short annealing times $t$. The solution used for fitting to all diffusion profiles is given by \cite{StrohmPhD}
\begin{equation}\label{eq:strohm}\begin{split}
&C \left( x,t\right) = \\
&\frac{C_0 / 2}{ \left({1 + \left(2Dt/ \sigma^2 \right)}\right)^{1/2}} \Biggl[ \rm{erfc} \left(\frac{ -\left(x_0 / 2 \sigma^2 \right) - \left( x/4Dt \right)}{ \sqrt{ \left( 1 / 2 \sigma^2 \right) + \left( 1 / 4Dt \right) }} \right) \\ 
& \exp{\left(\frac{-\left( x - x_0\right) ^2}{2 \sigma^2 + 4Dt} \right) } +k \,\rm{erfc} \left(\frac{ -\left(x_0 / 2 \sigma^2 \right) + \left( x/4Dt \right)}{ \sqrt{ \left( 1 / 2 \sigma^2 \right) + \left( 1 / 4Dt \right) }} \right) \\ 
& \exp{ \left(\frac{-\left( x + x_0\right) ^2}{2 \sigma^2 + 4Dt} \right)} \Biggl],
\end{split}\end{equation}
where $x$ is the depth normal to the surface, $C_0$ is the maximum surface concentration of the implantation profile and $D$ is the diffusion coefficient. The parameter $k$ accounts for the fact that the surface can act as a perfect reflector for the atoms $\left(k = +1 \right)$ or as a perfect sink $\left(k = -1 \right)$, or as a boundary with mixed reflector/sink properties $\left( -1 < k < 1\right) $. The derived diffusion coefficients are listed in table \ref{tab:table1} together with the corresponding parameters $x_{0}$ and $\sigma$ characterizing the initial tracer distribution. According to the concentration decline at the surface, the deduced $k$-values are as low as $ -0.95$ and the surface is almost a perfect sink for ${}^{43}\rm{K}$.
\begin{table}
\caption{\label{tab:table1}Diffusion coefficients of ${}^{43}\rm{K}$ in alkali feldspar normal to $\left( 001 \right)$ according to fitting of equation (\ref{eq:strohm}) to the diffusion profiles (cf. figure \ref{fig:2}). The initial tracer distribution after implantation is given by the Gaussian function (\ref{eq:gauss}) with the parameters $x_0$ and $\sigma$.}
\begin{tabular}{@{}llllll}
\hline\noalign{\smallskip}
$T$ (K) & $t$ (s) & $x_0$ (nm) & $\sigma$ (nm)& $k$ &$D \,\, \left( \rm{m}^2/\rm{s} \right) $ \\
\noalign{\smallskip}\hline\noalign{\smallskip}
1169 & 1244 & 50.7 & 19.7 & -0.96 & $1.9 \times 10^{-16}$ \\
1123 & 1847 & 50.7 & 19.7 & -0.94 & $1.0 \times 10^{-16}$ \\
1073 & 3795 & 50.7 & 19.7 & -0.95 & $4.2 \times 10^{-17}$ \\
1021 & 11,540 & 50.7 & 19.7 & -0.94 & $5.8 \times 10^{-18}$ \\
\noalign{\smallskip}\hline
\end{tabular}
\end{table}
The temperature dependence of self-diffusion is commonly described by the Arrhenius relation
\begin{equation}
D^\ast \left( T \right) = D_0 \exp{\left(-\frac{Q}{k_{\rm B} T}\right)},
\label{eq:arrhenius}
\end{equation}
where $D_0$ is the pre-exponential factor, $k_{\rm B}$ is the Boltzmann constant and $Q$ denotes the activation energy. In figure \ref{fig:3}, the resulting K self-diffusion coefficients $D_{\rm K}^\ast$ normal to $\left( 001 \right)$ are shown on a logarithmic scale as a function of $1/T$. It can be seen that the $D_{\rm{K}}^\ast$-values are fairly well described by the Arrhenius equation with $Q = \left( 2.4 \pm 0.4 \right) \,\, \rm eV$ and $\ln{D_{0} / \rm{m}^2 \rm{s}^{-1}} = -12.2 \pm 3.9$. The relatively large uncertainties in $Q$ and $D_0 = 5 \times 10^{-6} \,\, \rm m^2/\rm s$ relate to the fairly narrow temperature range $\left(1021-1169 \,\, \rm{K} \right)$, which is constrained both at high and low $T$ by the specific conditions of the experiments (short half-life time, lower limit of annealing time, surface acting as a sink). However, the present results for $D_{\rm K}^\ast$ are sufficiently reliable and accurate to compare with Na self-diffusion and ionic conductivity data pertaining to the same alkali feldspar and the same crystallographic direction, as will be done in the following sections.

\subsection{Comparison of Na and K self-diffusion}

In figure \ref{fig:3}, Na self-diffusion coefficients $D_{\rm Na}^{\ast}$ for Volkesfeld feldspar normal to $\left(001\right)$ from a previous study \cite{wilangowski15} are included for comparison. It is seen that $D_{\rm Na}^\ast \gg D_{\rm K}^{\ast}$ is fulfilled over the whole investigated temperature range and that, e.g., at $1173 \,\, \rm K$ the ratio $D_{\rm Na}^\ast / D_{\rm K}^\ast$ derived from the fitted Arrhenius equation is $1230 / 1$. Since both alkali components share the same sublattice, this finding suggests that the pertaining atomic jump frequencies are greatly different, i.e. $\wNa \gg \wK$. Under this assumption, point-defect-based diffusion mechanisms are subject to significant atomic correlation effects, especially when a K-rich feldspar such as VF is considered. This conclusion can be understood as follows: Below a certain Na site fraction referred to as percolation threshold $p_c$, long-range diffusion via the vacancy or the interstitialcy mechanism also requires atomic jumps via slow substitutional K atoms. The diffusion coefficient $D_{\rm{Na}}^\ast$ will then be coupled to $D_{\rm K}^\ast$ because both components share the same type of defect acting as a diffusion vehicle. Regardless of the ratio $\wNa / \wK$ this results in an upper limit for the ratio $D_{\rm Na}^{\ast} / D_{\rm K}^{\ast}$. For a vacancy-based diffusion model in alkali feldspar, a previous Monte Carlo (MC) study \cite{wilangowski15} determined this ratio to be $D_{\rm Na}^{\ast}/D_{\rm K}^{\ast} \leq 3.12$ for a VF-like K site fraction. This finding is in clear contrast with the experimental data presented in figure \ref{fig:3} and rules out the possibility that both Na and K diffusion are controlled by vacancies.

In a recent MC study \cite{wilangowski16}, correlation factors $f$ and the percolation threshold $p_c$ for an interstitialcy diffusion mechanism in a simple cubic reference structure with allowance for \emph{non}-colinear jumps of the Na ions have been determined. The findings for this reference system are partly transferable to the sublattice of alkali feldspar, as will be demonstrated in a forthcoming paper about an MC interstitialcy model and diffusion correlation effects in alkali feldspar, since both structures share the same number of neighbouring lattice sites to a self-interstitial. Particularly the percolation threshold seems unaffected by a transition from the cubic to the monoclinic system and is found to be $p_{c} = 0.122$, which is below the Na site fraction of VF $\left(C_{\rm Na} = 0.15 \right)$. This observation complies with $D_{\rm Na}^\ast / D_{\rm K}^\ast \geq 1000$ and the following discussion will therefore be focused on this diffusion mechanism.

\subsection{Concentration dependence of K self-diffusion}

${}^{41}\rm K$ diffusion experiments have been performed by Foland \cite{foland74} using a natural orthoclase from Benson Mines (BM), USA, with a K site fraction that is $C_{\rm{K}} \approx 0.94$ and therefore different from VF with $C_{\rm{K}}=0.83$. The activation energy pertaining to K self-diffusion was determined as $Q = 2.95 \,\, \rm{eV}$, which is somewhat above the value of $2.4 \,\, \rm{eV}$ obtained for VF in the present work. However, this difference is not significant because of the appreciable uncertainties in $Q$ both in our work $\left( \pm 0.4 \,\, \rm{eV} \right)$ and in the work of Foland ($\pm 0.2 \,\, \rm{eV}$, estimated by us from his reported data). A direct comparison between the K diffusion coefficients of both studies at $1073 \,\, \rm{K}$ (this is the only coincident temperature) gives $D_{\rm{K,VF}}^\ast / D_{\rm{K,BM}}^\ast \approx 1.8$ (cf. figure \ref{fig:3}). It should be noted that Foland's evaluation of $D_{\rm{K,BM}}^{\ast}$ is based on two different feldspar grain geometries, a \emph{spherical} and a \emph{cylindrical} one, leading to the relationship $D_{\rm{cylindrical}} \approx 2.2 \,\, D_{\rm{spherical}}$ (and hence $D_{\rm{K,VF}}^\ast / D_{\rm{K,BM}}^\ast \approx 4.0$ for the \emph{spherical} grain model). The following discussion is based on the values from the \emph{cylindrical} diffusion model. 

The good agreement between the two $D_{\rm{K}}^{\ast}$ values observed for the VF $\left(C_{\rm{K}}=0.83 \right)$ and BM $\left(C_{\rm{K}} \approx 0.94 \right)$ feldspars requires an examination of the concentration dependence of K self-diffusion. Two major effects are considered: Under the assumption of a non-colinear interstitialcy diffusion mechanism the concentration dependence of correlation factors $f_{\rm{K}}$ may be estimated from a Monte Carlo study \cite{wilangowski16}. The slower moving K ions are subject to correlation effects with a weak concentration dependence in the composition range from $C_{\rm{K}} = 0.83$ to $0.94$, and we find $f_{\rm{K,VF}} / f_{\rm{K,BM}} \approx 2 / 1$. A second effect on concentration dependence is the concentration of self-interstitials $C_{\rm{i}}$ acting as diffusion vehicles. In a previous study about ionic conductivity in feldspar \cite{hamid15} it was shown that the concentration of Na self-interstitials $C_{\rm{Na,i}}$ dominates over K self-interstitials $C_{\rm{K,i}}$ by orders of magnitude in the composition regime under discussion. We may therefore estimate $C_{\rm{i}} \approx \left[ C_{\rm{Na}} \exp{\left(G^{\rm{FP}}_{\rm{Na}} / k_{\rm{B}}T \right)} \right]^{1/2}$, where $G^{\rm{FP}}_{\rm{Na}}$ is a virtual $C_{\rm{Na}}$-independent free enthalpy of Frenkel pair formation and find $ C_{\rm{i}}^{\rm{VF}} / C_{\rm{i}}^{\rm{BM}} = \left( C_{\rm{Na}}^{\rm{VF}} / C_{\rm{Na}}^{\rm{BM}} \right)^{1/2} = \left(0.15 / 0.06 \right)^{1/2} \approx 1.6$. Combination of the two effects gives $D_{\rm{K,VF}}^\ast / D_{\rm{K,BM}}^\ast \approx 3$, which is in sufficient agreement with the experimental results when large uncertainties are considered. However we note that the composition of feldspar may have an effect on atomic potentials and hence on atomic jump frequencies $\omega$, which will not be considered here.

\subsection{Ionic conductivity derived from tracer diffusion}

\begin{figure}
\includegraphics[width=1.0\columnwidth ,clip]{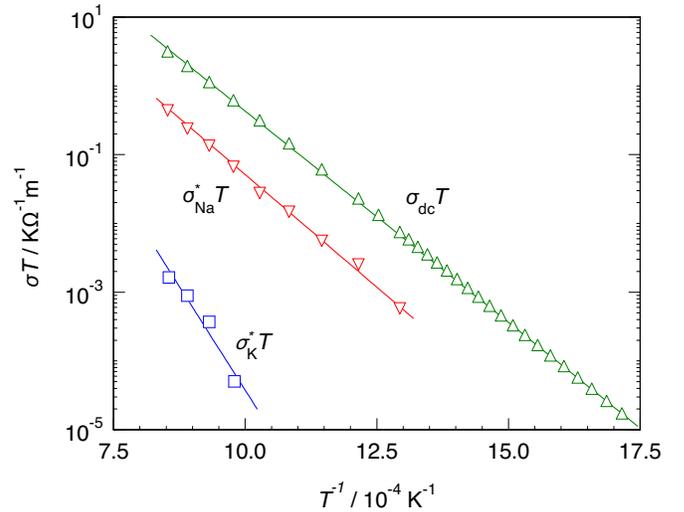}
\caption{Arrhenius plot of $\sigma_{\rm K}^{\ast} T$ derived according to (\ref{eq:sigma}) from diffusion coefficients. The results are compared with the partial conductivity $\sigma_{\rm Na}^{\ast} T$ \cite{wilangowski15} and with the total ionic conductivity $\sigma T$ obtained from impedance spectroscopy \cite{hamid15}.}
\label{fig:4}
\end{figure}

\begin{figure}
\includegraphics[width=1.0\columnwidth ,clip]{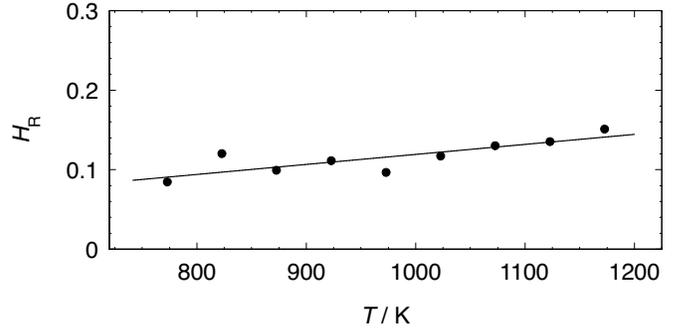}
\caption{Haven ratio $H_{\rm R}$ as a function of temperature $T$ derived according to equation (\ref{eq:haven}) from ionic conductivity data \cite{hamid15} and from Na self-diffusion results \cite{wilangowski15}.}
\label{fig:5}
\end{figure}

According to the Nernst-Einstein equation the self-diffusion coefficient $D_{\rm{K}}^{\ast}$ may be converted into an ionic conductivity $\sigma^{\ast}_{\rm{K}}$ \cite{wilangowski15, hamid15}, i.e., 
\begin{equation}
\sigma^{\ast}_{\rm{K}} T = \frac{e^2 N_{\rm{K}}}{k_{\rm{B}}} D_{\rm{K}}^{\ast},
\label{eq:sigma}
\end{equation}
where $e$ is the electronic charge of K and $N_{\rm K}$ denotes the concentration of K in ions per volume (number density). For VF feldspar with a measured density $\rho = 2.54 \,\, \rm g/\rm cm^{3}$, the K concentration is $N_{\rm{K}} = 4.67 \times 10^{21} \,\, \rm cm^{-3}$. Using a similar equation to (\ref{eq:sigma}) for Na, we insert the volume concentration of Na ions, i.e. $N_{\rm Na} = 8.01 \times 10^{20} \,\, \rm cm^{-3}$. The results derived for the partial Na and K conductivities $\sigma_{\rm Na}^{\ast}$ and $\sigma_{\rm K}^{\ast}$, respectively, are shown in figure \ref{fig:4} together with the total conductivity $\sigma$ measured by impedance spectroscopy \cite{hamid15}. It can be seen that although $N_{\rm Na} < N_{\rm K}$ holds true for the ion concentrations, the partial conductivity of Na clearly dominates over K over the whole temperature range, i.e. $\sigma_{\rm Na}^{\ast} \gg \sigma_{\rm K}^{\ast}$. The influence of $\sigma_{\rm K}^{\ast}$ on the total ionic conductivity may therefore be neglected, so that $\sigma^{\ast} = \sigma_{\rm K}^{\ast} + \sigma_{\rm Na}^{\ast} \approx \sigma_{\rm Na}^{\ast}$. From figure \ref{fig:4} it can be inferred that $\sigma^{\ast} < \sigma$, which can be expressed as $H_{\rm R} < 1$, where the Haven ratio is given by
\begin{equation}
H_{\rm R} = \sigma^{\ast} / \sigma.
\label{eq:haven}
\end{equation}
The results derived for $H_{\rm R}$ are shown in figure \ref{fig:5} as a function of $T$. It can be seen that $H_{\rm R} \approx 0.1$ and that $H_{\rm R}$ slightly increases with increasing $T$. Based on these findings we discuss the consequences for self-diffusion in alkali feldspar in the following.

A direct interstitial mechanism operating alone predicts a Haven ratio of $H_{\rm R} = 1$, which implies that charge and mass transport are equal \cite{DYRE1986139}. It should therefore be concluded that such a mechanism does not dominate the diffusion of Na atoms. Apparently, direct interstitial (I-I) jumps are much less frequent than interstitialcy (I-S/S-I) jumps for Na in VF feldspar. Consequently, these state-of-the-art results and considerations demand an explanation for $H_{\rm R} \approx 0.1$ within an interstitialcy diffusion model. It should be noted that the Haven ratio is intimately connected with correlation effects. For instance, for a hypothetical Na-K simple cubic (sub)lattice with random order and exclusively colinear interstitialcy jumps the Haven ratio is given by \cite{hamid15, wilan-jcp}
\begin{equation}
 H_{\rm R} = \frac{f_{\rm Na}}{2 f_{I}},
\label{eq:haven-b}
\end{equation}
provided that Na is much more mobile than K. Here, $f_{\rm Na}$ is the tracer correlation factor for Na and $f_{I}$ is the correlation factor of the self-interstitials entering the expression for the ionic conductivity \cite{murch-dyre89}.

In alkali feldspars we are dealing with a much more complex situation. Inspection of the monoclinic structure reveals at least four different (non-colinear) jump types for both Na and K, which are characterized by jump length and multiplicity. Our current work is therefore focused on the development of a suitable interstitialcy diffusion model in alkali feldspar and MC simulation of the corresponding correlation effects. This implies practical attempts to reproduce the $D_{\rm Na}^{\ast}/D_{\rm K}^{\ast}$ ratios and $H_{\rm R}$ values that are found experimentally. A successful link between simulation and experiment would give further information related to atomic jump frequencies and defect concentrations. Another challenging task is to reproduce by MC simulation the dependence of tracer diffusion and ionic conductivity on crystallographic orientation. Currently this work is in progress at our laboratory. 

\section{\label{sect:conclusions}Summary and conclusions}

In this paper self-diffusion of K in natural alkali feldspar from Volkesfeld, Germany normal to $\left(001\right)$ was examined by the radiotracer method after implantation with ${}^{43}\rm{K}$. It was shown that within the investigated temperature range, i.e. from $1169 \,\, \rm{K}$ to $1021 \,\, \rm{K}$, the diffusivity $D_{\rm{K}}^{\ast}$ is described by an Arrhenius equation with activation energy $Q = 2.4 \,\, \rm{eV}$ and the pre-exponential factor $D_0 = 5 \times 10^{-6} \,\, \rm{m}^2/\rm{s}$. The ${}^{43}\rm{K}$ diffusion coefficients were compared to those of ${}^{22}\rm{Na}$ in the same alkali feldspar and the same crystallographic direction. The diffusivity ratio $D_{\rm{Na}}^\ast / D_{\rm{K}}^\ast$ was shown to be $1230 / 1$ at $1173 \,\, \rm{K}$, which is in good agreement with previous observations. From this finding it can be concluded that the vacancy mechanism must be rejected for controlling alkali diffusion in alkali feldspar.

The reported $D_{\rm K,VF}^{\ast}$ results from this study agree fairly well with similar data $D_{\rm K,BM}^{\ast}$ obtained for a natural orthoclase with a higher K site fraction. Using ionic conductivity results from a previous study it was shown that $H_{\rm R} \approx 0.1$. A predominance of indirect interstitial jumps (I-S/S-I) over direct ones (I-I) points to correlated motion of Na and K via a common interstitialcy mechanism.

\begin{acknowledgments}
We acknowledge the support provided by the Federal Ministry of Education and Research (BMBF) through grants 05K13TSA and 05K16PGA for the use of implantation and diffusion equipment and infrastructure at ISOLDE/CERN. We are grateful to Karl Johnston and Juliana Schell for their help in performing ${}^{43}\rm K$ implantations and to Marina Mu\~noz Castro for contributing surface profile measurements. This research was further supported by the Deutsche Forschungsgemeinschaft through the grant STO 210/16-1.
\end{acknowledgments}

\section*{}
This manuscript has been published in \textit{Phys. Chem. Minerals} (2016). The final publication is available at Springer via http://dx.doi.org/10.1007/s00269-016-0862-1.
\bibliographystyle{aipnum4-1}
\bibliography{PCM16}

\end{document}